\documentclass[12pt]{iopart}
\usepackage{graphicx}
\begin{document}

\title[Character of Plasmons in small MNPs]{Which resonances in small metallic nanoparticles are plasmonic?}

\author{Emily Townsend and Garnett W Bryant}

\address{Joint Quantum Institute and Quantum Measurement Division, National Institute of Standards and Technology, Gaithersburg, MD 20899-8423}

\ead{etownsend@eml.cc, garnett.bryant@nist.gov}
\begin{abstract}

 We use time-dependent density functional theory to examine the character of various resonances corresponding to peaks in the optical response of small metallic nanoparticles.  Each resonance has both ``sloshing'' and ``inversion'' character.  The sloshing mode is an oscillation in the occupation of the shells nearest the Fermi energy, transferring charge back and forth from below the Fermi level to above it.  It results in oscillation in charge density near the surface of the particle.  Inversions monotonically move charge from occupied to unoccupied states, and result in oscillation in charge density in the core of the particle.  We also discuss the dependence of the optical response on the size of the simulation grid, noting that the character of resonances appears stable with respect to changes in simulation size, even though the details of the spectrum change. This makes a reliable characterization possible. We consider what characteristics are important in deciding that a resonance is plasmonic.

\end{abstract}

\pacs{1315, 9440T}
\vspace{2pc}

\section{Introduction}

Plasmons in metallic nanoparticles (MNPs) could potentially be used as the quantum excitations in quantum computing, communication and measurement schemes.  Nanoscale transmission of quantum excitations can occur as the energy and quantum information in a plasmon move between a MNP and other nanoscale quantum emitters. Therefore, we need to know what characterizes excitations in finite metallic systems that are small enough to be quantum, but large enough to have plasmon-like excitations. 

 The response of single atoms and small metallic clusters is described in terms of single electron transitions. 
In large particles, plasmons are classical oscillations of charge near the metal surface.  A clear characterization of excitations in intermediate size particles, small enough to have quantized single-particle excitations but large enough to sustain collective charge oscillations, is more difficult.  This question has been discussed for many years.

 Theoretical quantum descriptions in infinite or semi-infinite systems can provide guidance but clearly don't describe MNPs.  Models that help us understand the quantum properties of plasmons in finite systems must be employed.   Classical hydrodynamic descriptions won't capture size quantization, quantum (coherent) coupling and quantum hybridization of states in multiple nanoscale particles, or interparticle tunneling. 
 Quantum electronic structure calculations have been used for thirty years to describe small metal particles.  Typically density functional theory (DFT)and time-dependent density functional theory (TDDFT) have been used.
\cite{Ekardt1985I, PuskaNieminenManninen1985, Beck1987,YannouleasBroglia1991a, Brack1993,VasilievOgutChelikowsky1999, YanYuanGao2007, AikensLiSchatz2008, YanGao2008,KummelAndraeReinhard2001,Quijada2010,MortonSilversteinJensen2011,RaitzaReinholzReinhardRopkeBroda2012,BernadotteEversJacob2013, PicciniHavenithBroerStener2013,BurgessKeast2014, TownsendBryant2012} 

\begin{figure}
 \includegraphics[width=0.65\textwidth]{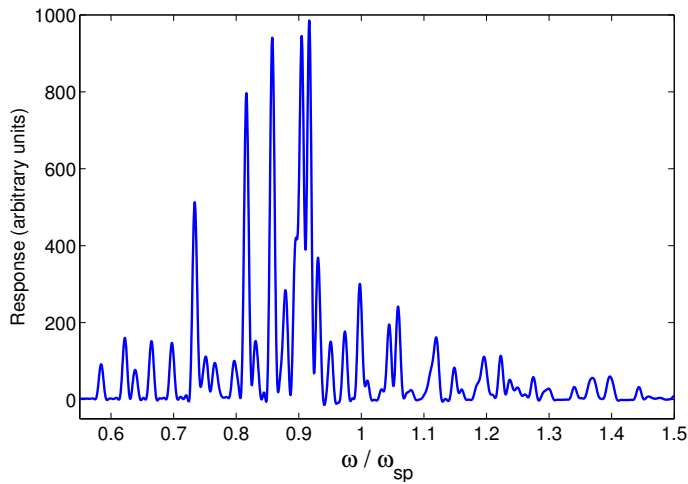}
 \caption{Optical response of a 100 electron MNP.  Horizontal axis measured in units of the classical surface plasmon frequency.} 
 \label{fig:deltaResponse}
 \end{figure}

 A jellium model for the ion cores allows a quantum mechanical description of delocalized interacting valence electrons that captures the important elements described above \cite{Brack1993}.  TDDFT can be used to calculate linear response functions or determine the time evolution of the response.\cite{TownsendBryant2012}   Figure \ref{fig:deltaResponse} shows an example of the optical response for a 100 electron jellium sphere calculated using real-time TDDFT.  (The horizontal axis is measured in units of the classical surface plasmon resonance, $\omega_{\rm sp} = \omega_{\rm p}/\sqrt 3 = [(n_e e^2)/(3m\varepsilon_o)]^{1/2}$, where $\omega_{\rm p}$ is the bulk plasma frequency, $n_e$ is the electron density, $e$ and $m$ are the electronic charge and mass and $\varepsilon_o$ is the vacuum permittivity.) 

Ekardt \cite{Ekardt1985I}, Puska, et al. \cite{PuskaNieminenManninen1985}, and Beck \cite{Beck1987} were some of the first authors to use frequency-space (linear response) TDDFT on jellium spheres to calculate the optical response of MNPs.  The response functions they calculated are similar to the response shown in \ref{fig:deltaResponse}, though with less fine frequency resolution. We expect that the different peaks seen in the optical response have different character, and different authors have attempted to tease out what this character might be, and which peaks might best be described as surface plasmons.  Generally the spectrum is described as one broad surface plasmon peak that is fragmented into multiple peaks by various interactions, plus many additional smaller peaks.  (There is also a second, much smaller, broad and fragmented peak at slightly higher energy that is visible if the amplitude is plotted on a logarithmic scale. This is generally identified as a volume plasmon.)

Beck \cite{Beck1987} looked at the induced electron density of the eigenmodes corresponding to some of these peaks and characterized those with induced charge near the surface of the particle as collective excitations and those with induced charge near the core of the particle as single-particle  excitations (particle-hole transitions).  However he cited trouble making these identifications clearly due to coupling of single-particle and collective excitations.  He stated that a collective or plasma excitation is one where the electronic charge density as a whole is excited and oscillates, and ``none of the electrons are transferred from an occupied to an unoccupied energy level and the induced density ... can be an infinitesimal perturbation of the ground-state system.''  

Ekardt \cite{Ekardt1985I} identified ``fine cusps'' in the spectrum to be single-particle-hole pairs and the broader central feature as a Mie resonance but did not identify any fragmentation. He examined the induced electron density in the radial direction as a function of frequency for some of these eigenmodes.  From this he noted that one of the eigenmodes he identified as a single particle-hole excitation had very little change in the charge density near the surface, with more change in the interior of the sphere.  The opposite was true for the eigenmode he identified as the surface resonance.
 
Puska, et al. \cite{PuskaNieminenManninen1985} also identified bound-bound cusps and bound-continuum edges in their spectra.  Their calculations were done both with and without electron-electron interactions.  Broader features were present only when electron-electron interactions are turned on, whereas there were narrow cusp-like peaks both with and without electron-electron interaction.

In a review article, Brack \cite{Brack1993} attributed the fragmentation of the surface plasmon peak to coupling between the surface plasmon and particle-hole excitations, volume plasmons, or more complicated excitations.

Yannouleas et al. \cite{YannouleasBroglia1991a}, using discrete-matrix random-phase approximation (RPA) of jellium spheres, calculated RPA oscillator-strength functions and showed that the surface plasmon is ``fragmented'' by a renormalization (shift) of the plasmon energy due to single electron transitions with change in principle quantum number of 1, followed by hybridization with the rest of the single electron transitions, some of which are very near the renormalized surface plasmon.  

Kummel et al. \cite{KummelAndraeReinhard2001} identified plasmon-like behavior in Na$_2$ based on the fact that local current approximation quantum fluid dynamics provided a description very similar to TDDFT, suggesting a convergence between viewing the excitations as valence electron oscillations and as transitions between molecular states.

Bernadotte et al. \cite{BernadotteEversJacob2013} identified plasmons in molecules by scaling the electron-electron interaction by a parameter $\lambda$, in TDDFT.  Single particle excitations have little or no dependence on $\lambda$, while plasmon excitations scale linearly with $\lambda$.
Piccini et al. \cite{PicciniHavenithBroerStener2013} simplified the analysis of Bernadotte et al. Rather than a full scaling with multiple values of $\lambda$, they examined the full TDDFT and Kohn-Sham transitions which are the $\lambda=1$ and $\lambda=0$ limits of the scaling.  
Raitza et al \cite{RaitzaReinholzReinhardRopkeBroda2012} identified whether excitations were collective (plasmonic) based on a bi-local auto-correlation function.

Previously, we have used real-space and real-time TDDFT to examine the spatial and time variation of the various eigenmodes of these jellium spheres \cite{TownsendBryant2012}.  We have seen that some eigenmodes have their induced charge density fluctuations near the surface of the sphere, referred to as ``classical surface plasmons'', and some in the interior, called ``quantum core plasmons''.  In \cite{TownsendBryant2012} we identified both of these kinds of eigenmodes as collective states, because electrons from multiple shells were contributing to the change in charge density in both cases.  

In the present article we continue this investigation into the character of the different resonances present in single MNPs. Using real-space, real-time TDDFT, we examine the transitions of electrons between different Kohn-Sham states as a function of time, and identify two types of behavior that each resonance exhibits.  Each resonant mode has electrons sloshing back and forth between shells above and below the Fermi surface and excitation of electrons continuously from occupied states to unoccupied states.  We consider the possible interpretations of these two types of transitions in identifying collective plasmon resonances in small MNP systems.  We also discuss the unexpected dependence of our absorption spectra on the size of our (spherical) simulation box.   The TDDFT simulations in the current work are refinements of those in \cite{TownsendBryant2012}.  New to this work is the analysis based on the projections of time-dependent single particle states onto single particle ground states.  This new analysis shows the presence of both sloshing and inversion modes in each of the classical surface plasmons and quantum core plasmons.  Many new calculations have also been done varying the size of the simulation boundaries.

\section{Methods} 
For the TDDFT calculations we use Octopus, a Gnu-Public Licensed software package \cite{Octopus2003, Octopus2006}.  The MNP is modelled as a gold jellium sphere  with  single-electron radius $r_s = 3 \, a_o = 0.159 \,\rm{nm}$.  We consider particles with  radius of $R_{\rm MNP} = 8.42\, a_o = 0.446 \,\rm{nm} $ (20 valence elctrons)  to $R_{\rm MNP} = 13.9\, a_o = 0.737\, {\rm nm}$ (100 valence electrons).  The particle is defined on a real-space grid (12 to 16 grid points per $R_{\rm MNP}$). The grid is defined inside a simulation sphere with a radius of 2 to 6 times the size of the particle.  The charge density and wave functions must go to zero at the edges of the simulation sphere.   The ground state of the jellium sphere is calculated self-consistently in the Local Density Approximation (LDA).  This gives the Kohn-Sham single particle eigenstates. 
In our DFT calculations, the electron density is created by filling the Kohn-Sham states with electrons according to a finite-temperature Fermi function, $f_n = 1/(\exp (E_n/k_{\rm B}T_{\rm smearing}) +1)$, where $E_n$ is the energy of a Kohn-Sham state, $k_{\rm B}$ is the Boltmann constant, and $T_{\rm smearing}$ is a temperature parameter that determines the smearing of levels.  Typically we are using an electronic temperature of 0.001 $E_h/k_{\rm B}$, which is 0.005 $\omega_{\rm sp}$ or 316 K. $E_h$ is a hartree, the energy unit used by Octopus.

We first do a ground state calculation, then we evolve the Kohn-Sham states in time using an approximated, enforced time reversal symmetry propagator with a timestep of $0.1 \, \hbar/E_{\rm h} = 1.5 \times 10^{-8} \, \rm{ns}$.  This approximates applying the operator $\exp(-i\hat Ht/\hbar)$ to the Kohn-Sham states in small time steps.  The exponential is Taylor expanded to fourth order.

There are two different time-evolution calculations we do: the first finds the optical response as a function of frequency, the second characterizes each resonant mode.  In the first calculation a linearly-polarized delta pulse electromagnetic field is applied at $t=0$ and the states are propagated forward with no further applied potential.  The Fourier transform of the resulting time-dependent dipole moment gives the dynamical polarizability as a function of frequency, the imaginary part of which is proportional to the dipole strength function and the optical cross section, as seen in figure \ref{fig:deltaResponse}. The peaks in this response are the resonant frequencies of the system.  Long time evolutions improve the frequency resolution.  Up to the approximations of our calculation (Local Density Approximation and the truncation of the expansion of the Hamiltonian for the time evolution) this response is proportional to the density-density many-body Green's function which has peaks at the energies of all quasiparticle-hole pair excitations (renormalized single particle states) and collective modes (plasmons).

In the  second time-evolution calculation, we apply a sinusoidally-varying electric field oscillating at one of the resonant frequencies, driving the particle. This allows us to to examine the electron density, the Kohn-Sham eigenstates, and the dipole or higher order-multipole moments of the system as a function of space and time.  We can also project the time-evolved Kohn-Sham eigenstates onto the ground-state Kohn-Sham eigenstates to identify the single-particle transitions involved.

\section{Results}

\subsection{Transitions between Kohn-Sham states as a function of time and frequency}
We begin by examining the single-particle transitions that contribute to the resonances of a 100 electron MNP.  This will allow us to understand how the individual electrons come to behave collectively.

Figure \ref{fig:deltaResponse} shows the photoabsorption cross section of the 100 electron MNP when the radius of the simulation volume is 3.4 times the radius of the particle ($R_{\rm grid} = 3.4 R_{\rm MNP}$). There are five principal resonances, with a pair of very strong resonances at 0.9041 $\omega_{\rm sp}$ and 0.9124 $\omega_{\rm sp}$.  
  These two have a very similar character to one another, and exhibit charge oscillation throughout the particle, with significant charge oscillation near the surface.  This can be seen in figure \ref{fig:SurfaceDensity}, which shows the induced change in electron density along the driving axis of the 100 electron MNP when it is driven with an oscillating electric field at 0.9041 $\omega_{\rm sp}$
, the frequency of the strongest peak in figure \ref{fig:deltaResponse}. 
  The horizontal axis is time measured in  the period of the driving frequency,  $T=2\pi/\omega_{\rm driving}$ and the vertical axis is position along the axis measured in terms of the radius of the MNP, $R_{\rm MNP}$.  The dark dotted line indicates the ground-state electron density as a function of position along the y-axis.
These resonances are described in reference \cite{TownsendBryant2012} as classical surface plasmons  (these two peaks were not resolved as separate peaks in that paper).  There are two moderately strong resonances at 0.8163 $\omega_{\rm sp}$ and  0.7337 $\omega_{\rm sp}$. 
  These are the quantum core plasmons of reference \cite{TownsendBryant2012}, in which the charge oscillation occurs near the center of the MNP.  There is an additional strong resonance at 0.8579 $\omega_{\rm sp}$
which has mixed character, with charge oscillations at the surface of the particle for some times in the evolution, but not consistently throughout the time evolution. 
All of the resonances are made from multiple single-particle transitions.\cite{TownsendBryant2012}

To understand the different behavior of these resonances, we must understand how electrons in different shells behave. For this purpose we project the time-dependent Kohn-Sham states, $\phi_n(t)$, onto the ground-state ($t=0$) Kohn-Sham states, $\phi_m(0)$, where $n$ and $m$ are state indices ordered by energy ($E_n$).  As the $n^{\rm th}$ state evolves in time its occupation remains the same, however it acquires components from states other than $n$ in the ground-state basis.   Each evolving state remains mostly in its original ($t=0$) Kohn-Sham state, but evolves with time to contain other ground-state ($t=0$) components.
Thus there is some probability that the electron has made a transition from the $n^{th}$ Kohn-Sham state to the $m^{th}$ Kohn-Sham state.  The square of the projection $<\phi_n(t)|\phi_m(0)>$ gives the probability that at time, $t$, the $n^{\rm th}$  eigenstate will have evolved into the $m^{\rm th}$ ground-state eigenstate.

Because we are interested in transitions, we want to know how much an occupied evolving state remains in or moves out of the Fermi sea. 
 The probability that any occupied state has evolved into the $m^{\rm th}$ ground-state Kohn-Sham state is:
\begin{equation}
P(m,t) = \sum_n f_n |<\phi_n(t)|\phi_m(0)>|^2 .
\end{equation}
The driving field amplitude is small, so very little charge is moving, and the projections are nearly a delta function, $\delta_{nm}$.  To see the variation, we instead plot the change in this probability:
\begin{equation}
\Delta P(m,t) = \sum_n f_n |<\phi_n(t)|\phi_m(0)>|^2  - f_m.
\end{equation}
The occupation of each evolving Kohn-Sham state is constant.  $P(m,t)$ shows how an occupied state evolves into another $t=0$ state.  For simplicity, however, we will sometimes refer to $P(m,t)$ as occupation and $\Delta P(m,t)$ as a change in occupation of the Kohn-Sham states.

By examining the time- and frequency-dependence of the transitions between Kohn-Sham states, we can identify which shells are participating in a resonance and how.  

\subsubsection{Classical Surface Plasmon}
\begin{figure}
  \includegraphics[width=0.55\textwidth]{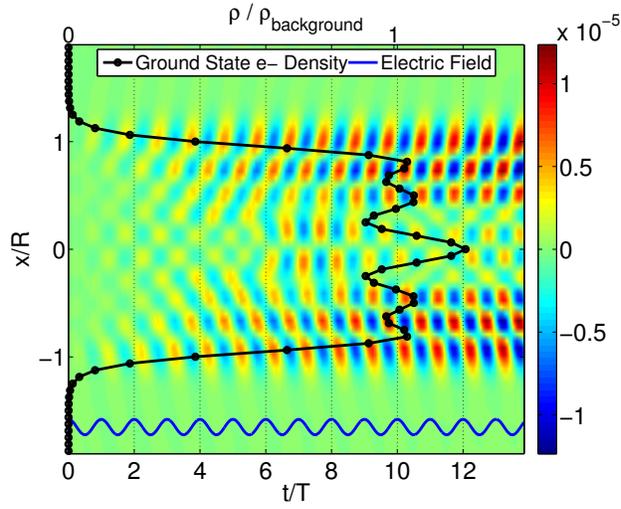}

 \caption{Classical surface plasmon ($\omega = 0.9041 \, \omega_{\rm sp}$) in 100 electron MNP: color indicates change in the electron density along the axis ($x$) parallel to the applied field through the center of the MNP.  Bottom blue line indicates the phase of the applied electric field as a function of time. Solid line with circles is the ground state electron density as a function of position, whose horizontal axis is at the top of the figure.  T is the period of the driving field.}
  \label{fig:SurfaceDensity}
 \end{figure}

\begin{figure}
  \includegraphics[width=0.55\textwidth]{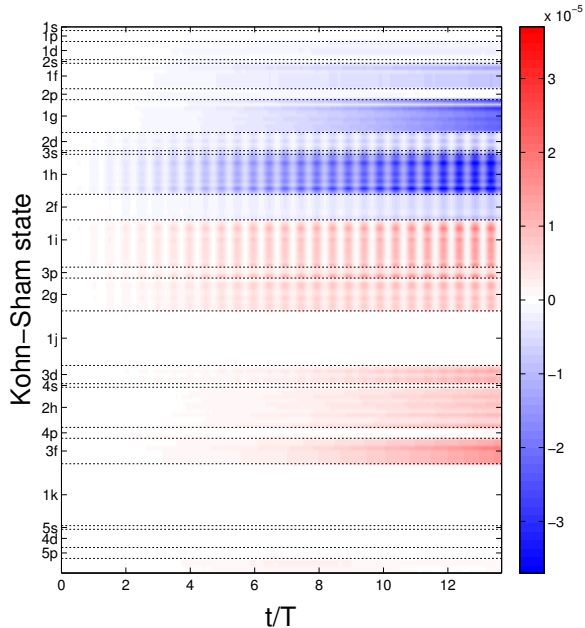}
 \caption{Change in the probability of occupation of Kohn-Sham  states ($ \Delta P(m,t) = \sum_n f_n |<\phi_n(t)|\phi_m(0)>|^2 -f_m $) for the classical surface plasmon in the 100 electron MNP.  The Fermi energy is in the middle of the 2f shell. }
  \label{fig:SurfaceTimeOccup}
 \end{figure}

As mentioned above, the classical surface plasmon resonance exhibits charge oscillation at the surface of or throughout the particle.   This is apparent from figure \ref{fig:SurfaceDensity}. 
 Figure \ref{fig:SurfaceTimeOccup} shows  $\Delta P(m,t)$ for the 100 electron MNP driven at this resonance.  The positive (red) areas indicate shells that electrons have moved to, while negative (blue) areas indicate shells that electrons have left. 
The vertical axis identifies which shell each time-dependent Kohn-Sham state belongs to. (Shell identities are deduced from the degeneracy of the states.)
This figure shows two kinds of changes in the occupation of the Kohn-Sham states.  Shells near the Fermi level in the middle of 2f (shells 2d, 3s, 1h, 2f, 1i, 3p, and 2g) exhibit an oscillation in their occupation, while shells 1d, 1f, 1g, 3d, 2h and 3f exhibit a monotonic increase or decrease in their occupation. We identify the oscillating occupation as a ``sloshing'' of charge back and forth with the applied field, created by electrons near the Fermi surface.  The monotonic evolution we identify as an inversion, electrons deeper in the core being excited to higher energy states.  In all five of the (different frequency) resonances we have examined in this system, sloshing oscillatory occupation occurs between states just below and just above the Fermi energy.  The sloshing component corresponds closest to a plasmonic response with the charge near the Fermi level involved.  Different resonances have different states involved in the inversions.  

\begin{figure}
  \includegraphics[width=0.85\textwidth]{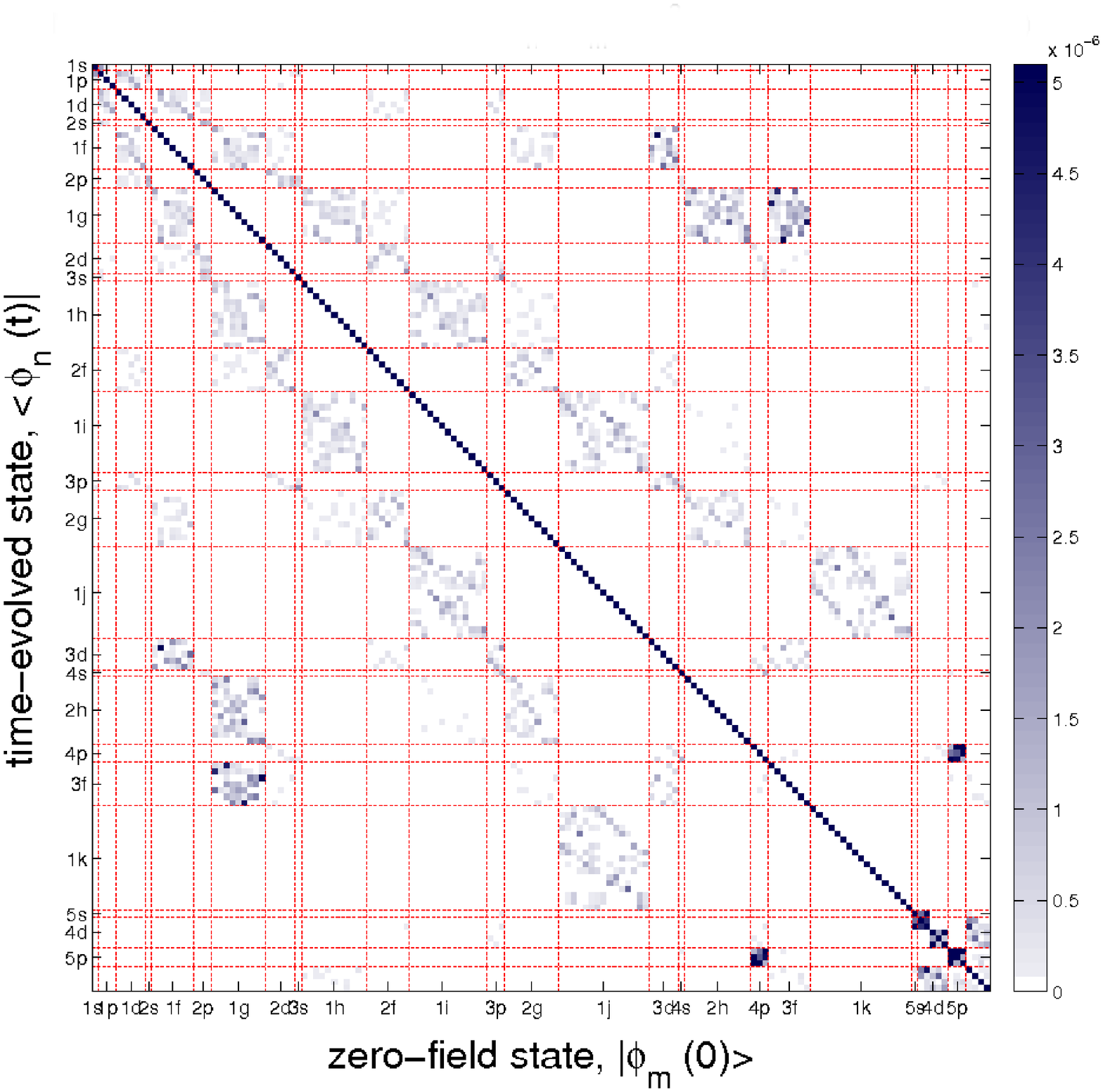}
 \caption{Magnitude Squared Projections ($|<\phi_n(t)|\phi_m(0)>|^2$) of time-evolved Kohn-Sham states onto the ground-state Kohn-Sham states for the classical surface plasmon in the 100 electron MNP. The snapshot is shown for an arbitrarily chosen time.}
  \label{fig:SurfaceProjections}
 \end{figure}

  By examining a snapshot of the projections of time-evolved Kohn-Sham states onto ground states (figure \ref{fig:SurfaceProjections}) we can confirm that the oscillations in charge leaving some shells are connected by transitions (non-zero projections) to oscillations in charge entering other shells (e.g. 2d $\rightarrow$ 3p (weak), 3s $\rightarrow$ 3p (moderate), 1h $\rightarrow$ 1i (moderate), 1h $\rightarrow$ 2g (weak), 1f $\rightarrow$ 3d (weak)).  Likewise, there are transitions between monotonically decreasing states and monotonically increasing states (e.g. 1f $\rightarrow$ 3d, 1g $\rightarrow$ 2h,  1g $\rightarrow$ 3f). 

\begin{figure}
  \includegraphics[width=0.55\textwidth]{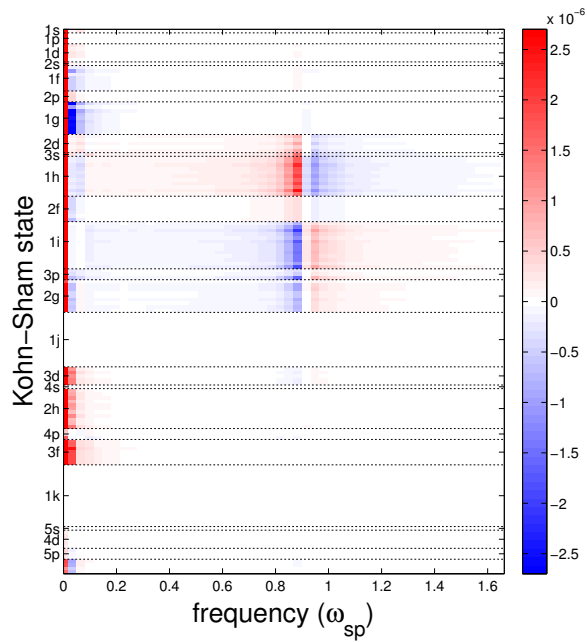}
  \caption{Real part of the Fourier transform of $\Delta P(m,t)$ for the classical surface plasmon in the 100 electron MNP.}
  \label{fig:SurfaceRealFreqOccup}
\end{figure}

\begin{figure}
  \includegraphics[width=0.55\textwidth]{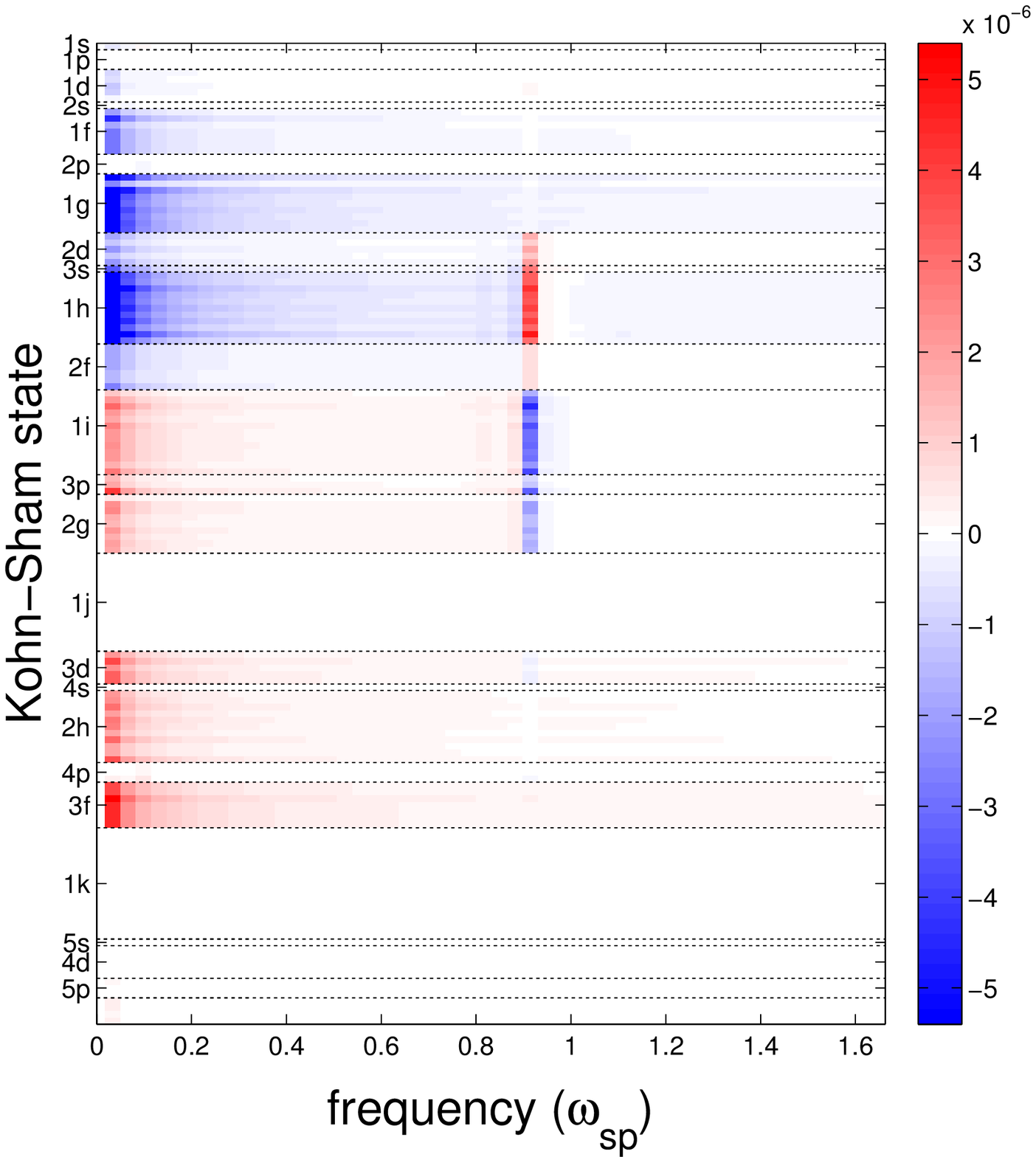}
  \caption{Imaginary part of the Fourier transform of $\Delta P(m,t)$ for the classical surface plasmon in the 100 electron MNP. }
  \label{fig:SurfaceImFreqOccup}
\end{figure}

Figures \ref{fig:SurfaceRealFreqOccup} and \ref{fig:SurfaceImFreqOccup} show the real and imaginary parts of the Fourier transform of figure \ref{fig:SurfaceTimeOccup}, showing the frequency behavior of the occupation of the shells.
We see a main peak consistent with the real and imaginary parts of the form $e^{i\phi}/(\omega - \omega_{\rm driving} - i\delta)$.  The oscillations in the occupation (sloshing) occur as peaks at the driving frequency.  The monotonic transitions (inversion) show up as peaks near zero frequency that extend broadly across many frequencies. We can also see that the monotonic inversion is also present in the shells with the oscillatory transitions. 

\subsubsection{Quantum Core Plasmons}

\begin{figure}
  \includegraphics[width=0.55\textwidth]{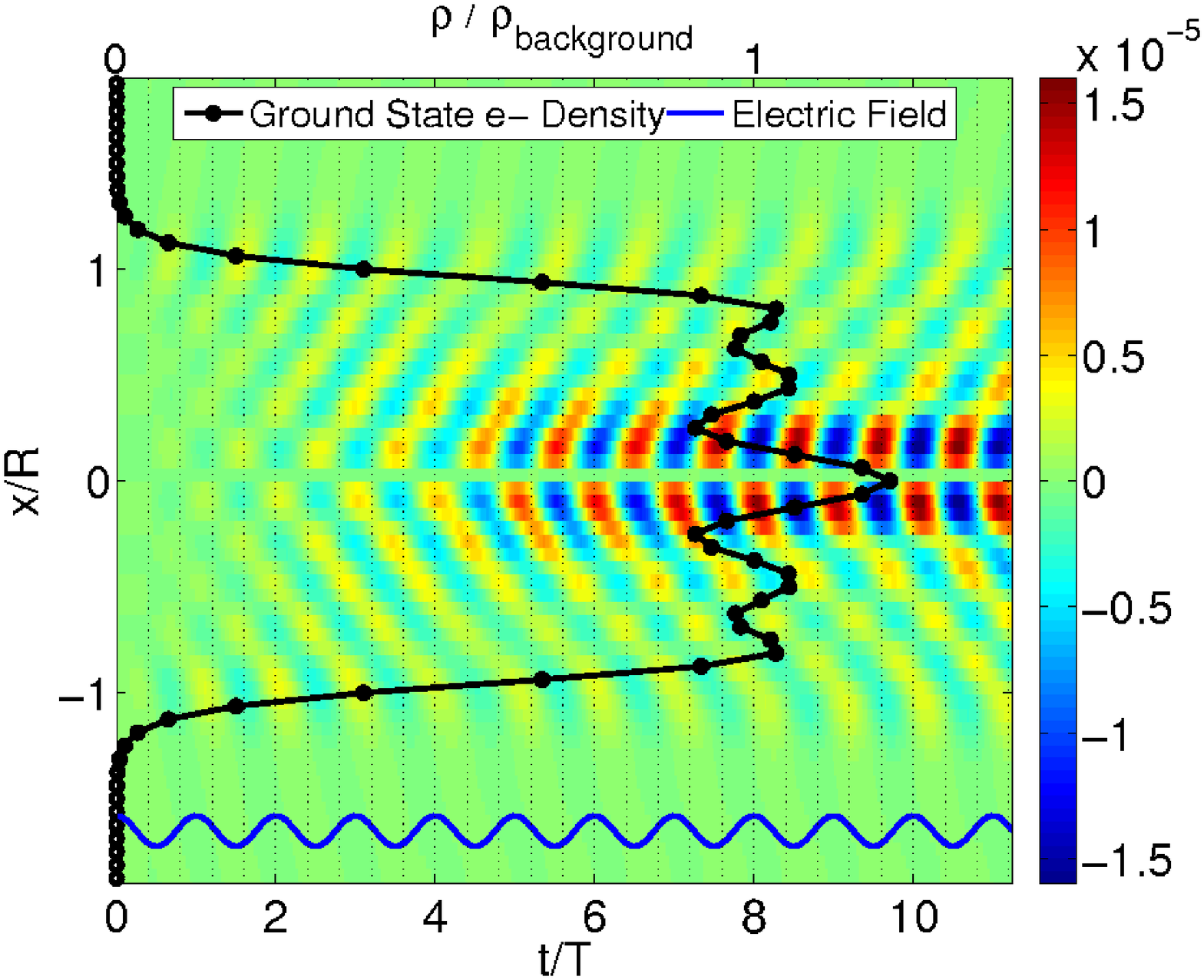}
 \caption{Quantum core plasmon ($\omega = 0.7337 \, \omega_{\rm sp}$) in the 100 electron MNP: color indicates change in electron density along the axis parallel to the applied field through the center of the MNP.  Bottom blue line indicates the phase of the applied electric field. Solid line with circles is the ground state charge density, whose horizontal axis is at the top of the figure.}
  \label{fig:CoreDensity}
 \end{figure}

\begin{figure}
  \includegraphics[width=0.55\textwidth]{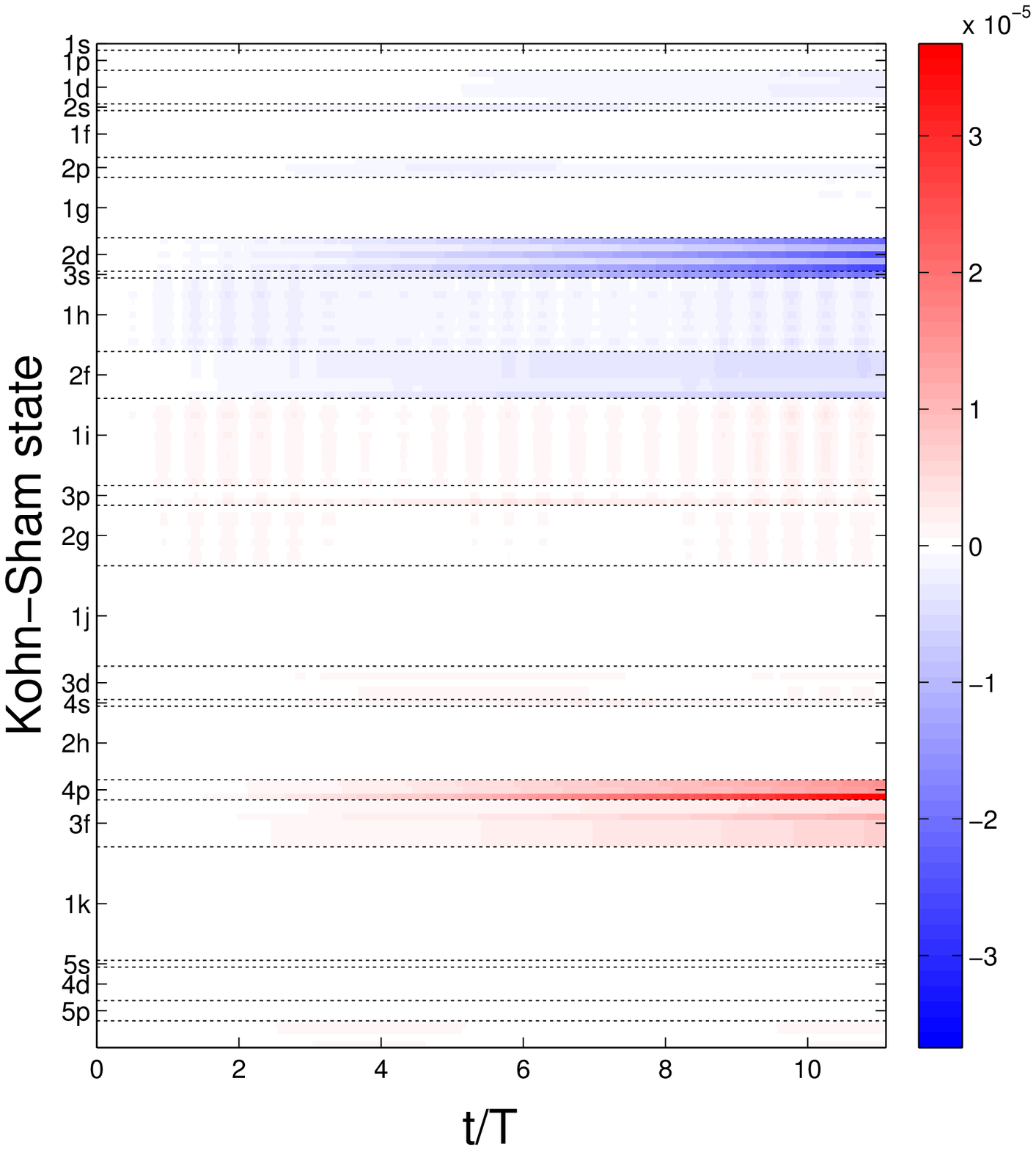}
 \caption{Change in occupation of Kohn-Sham  states ($ \Delta P(m,t) = \sum_m f_m |<\phi_n(t)|\phi_m(0)>|^2 $) for classical surface plasmon  in the 100 electron MNP.  The Fermi energy is in the middle of the 2f shell. }
  \label{fig:CoreTimeOccup}
 \end{figure}

\begin{figure}
  \includegraphics[width=0.85\textwidth]{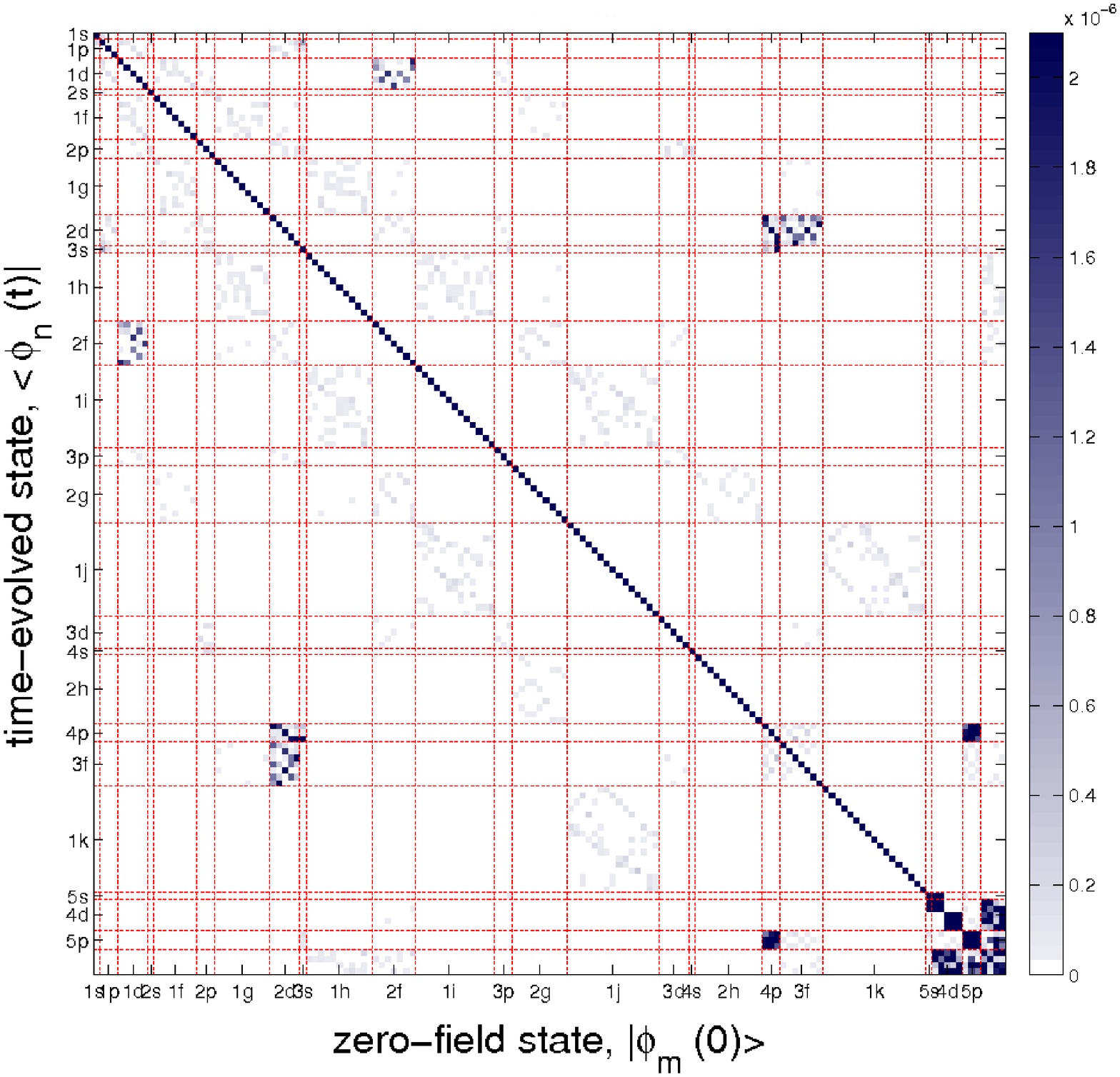}
 \caption{Magnitude Squared Projections ($ |<\phi_n(t)|\phi_m(0)>|^2$) of time-evolved Kohn-Sham states onto the ground-state Kohn-Sham states for the quantum core plasmon in the 100 electron MNP. Inset shows the indicated region with a color scale that is 20 times more sensitive.}
  \label{fig:CoreProjections}
 \end{figure}

\begin{figure}
  \includegraphics[width=0.65\textwidth]{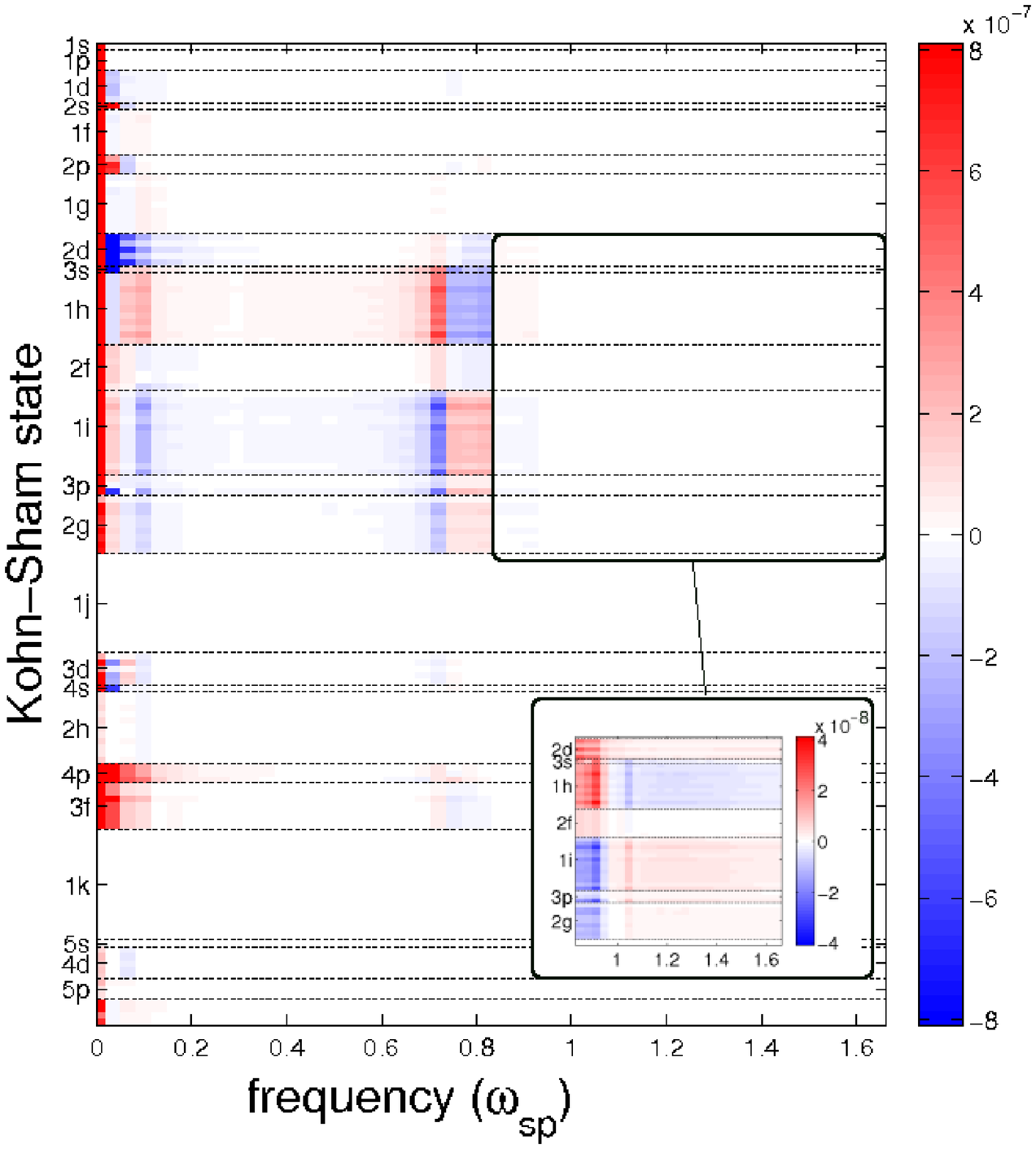}
  \caption{Real part of the Fourier transform of $\Delta P(m,t)$ for the classical surface plasmon in the 100 electron MNP. Inset shows the indicated region with a color scale that is 20 times more sensitive, to reveal additional weak peaks. }
  \label{fig:CoreRealFreqOccup}
 \end{figure}

\begin{figure}
  \includegraphics[width=0.65\textwidth]{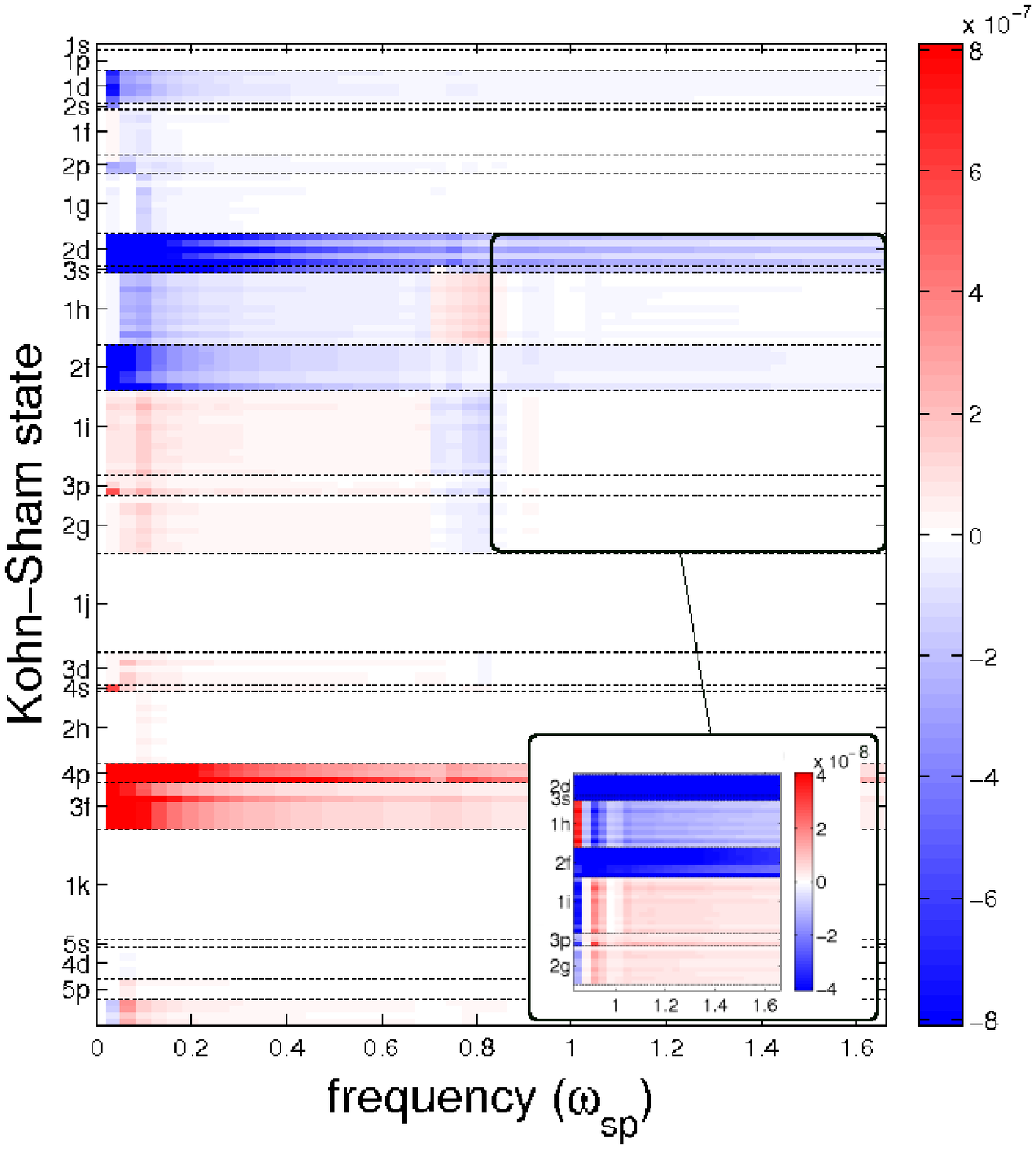}
  \caption{Imaginary part of the Fourier transform of $\Delta P(m,t)$ for the classical surface plasmon in the 100 electron MNP.  Inset shows the indicated region with a color scale that is 20 times more sensitive, to reveal additional weak peaks.}
  \label{fig:CoreImFreqOccup}
 \end{figure}

The corresponding results for one of the quantum core plasmon resonances  ($\omega = 0.7337 \, \omega_{\rm sp}$), with charge oscillation only in the MNP interior, are shown in figures \ref{fig:CoreDensity}, \ref{fig:CoreTimeOccup}, \ref{fig:CoreProjections}, \ref{fig:CoreRealFreqOccup} and \ref{fig:CoreImFreqOccup}.
 This resonance has only a few of the monotonic inversion transitions (3s $\rightarrow$ 4p, 2d $\rightarrow$ 4p, 2d $\rightarrow$ 3f) and the sloshing transitions are much weaker (1h $\rightarrow$ 1i, 2f $\rightarrow$ 2g, 2d $\rightarrow$ 4p (very weak), 2d $\rightarrow$ 3f (very weak), 2p $\rightarrow$ 3d (very weak), 3s $\rightarrow$ 4p (very weak), 2d $\rightarrow$ 4p(very weak)).  
States from different shells are involved in the inversions for the other core resonances.  
The insets in Figures \ref{fig:CoreRealFreqOccup} and \ref{fig:CoreImFreqOccup} show that the quantum core plasmon has a weak resonance at the classical surface plasmon frequency ($\omega = 0.9041 \, \omega_{\rm sp}$) in addition to its response at the driving frequency.

In all of the core resonances there is a response at both the driving frequency and, to some extent, at the frequency of the surface resonance. 
Also,  the relative strength of the peaks at $\omega = 0$ and $\omega=\omega_{\rm driving}$ are different for the core resonances and the surface resonances.  The core resonances have less to much less strength in the peak at the driving frequency.  The sloshing behavior is more dominant in the suface resonances and the inversion behavior in the core resonances.  This is consistent with the picture of a sloshing resonance being split or fragmented by transitions from core states into unoccupied states and the core resonances being more single-particle like.

Some states have mixed character and are not clearly either classical surface plasmons or quantum core plasmons.  For example, the mixed character state at 0.8579 $\omega_{\rm sp}$ has strong responses both at at $\omega = 0$ and $\omega=\omega_{\rm driving}$ with nearly all of the shells involved in the inversion transitions, but with one particular shell (1h) being the source of most of the inversion electrons.

\subsection{Variation with simulation volume}
To ensure that these assignments are robust to the details of the calculation, we have examined the dominant resonant modes of dozens of single MNPs, varying their size and the simulation parameters.  
The calculated photoabsorption cross sections are robust to variations in the magnitude of the delta kick, the length of the time step, the grid spacing, and the convergence energy for the self-consistency, indicating that we are in a regime of linear response and that our time step, grid spacing, and convergence energy are sufficiently small.  Because we Fourier transform a finite time series, we get better resolution in the photoabsorption as a function of frequency for longer time series. 

\begin{figure}
 \includegraphics[width=0.65\textwidth]{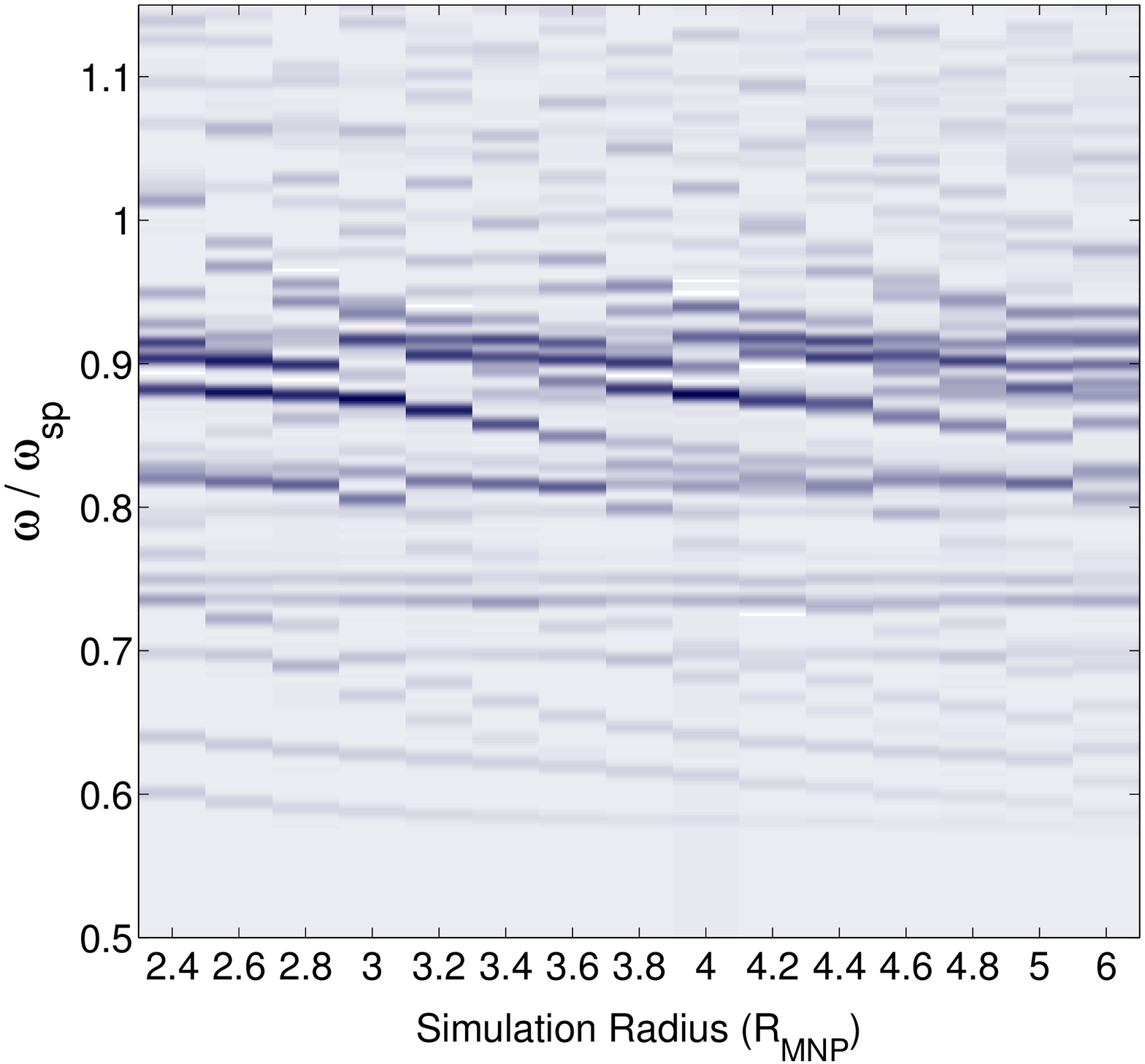}
 \caption{Photoabsorption cross section of 100 electron MNP, varying simulation size.  Darker colors indicate greater absorption, in arbitrary units.}
 \label{fig:CSVectorVaryBC}
 \end{figure}

For longer time series that provide better frequency resolution, a significant dependence of photoabsorption on the size of our simulation volume (a sphere)  becomes apparent, as seen in figure \ref{fig:CSVectorVaryBC}.  Color indicates the photoabsorption cross section, which is a function of frequency on the y-axis and a function of the radius of the simulation volume, $R_{\rm grid}$, in terms of the radius of the MNP, $R_{\rm MNP}$, on the x-axis.

Figure \ref{fig:CSVectorVaryBC} shows that the dominant spectral weight of the response remains at nearly a constant energy for a wide range of simulation sizes.  However the details of the ``fragmentation'' discussed in the introduction vary considerably.  There are always multiple peaks clustered around the same energy, but their location and relative strengths vary with $R_{\rm grid}$. Less prominent peaks appear to red shift with increasing volume, and it appears that these weaker peaks pass through the plasmon resonance, causing splittings. 
Even with an $R_{\rm grid}$ of six times the particle radius, the levels have not fully converged to stable positions or relative strengths.  

 Only a small fraction of the total charge is moving.  The maximum electron density in the particle is about 1000 times the maximum change in electron density in the particle.  An even smaller fraction of the charge is near the edges of the simulation grid.  The maximum electron density in the particle is at least $~10^9$ times the maximum electron density at the edge of the simulation during the entire time evolution. 
 Thus the charge density is not reaching the edges of the simulation volume in significant amounts. 
Hence, the bound, occupied Kohn-Sham states are not sensitive to the simulation size.  
 This suggests that unbound, continuum states are playing an important role in our calculation. The Fermi energy is at $-0.5702 \, \omega_{\rm sp}$ 
in the 100 electron MNP.  Since the primary resonances are in the range of 0.62  $\omega_{\rm sp}$ to 1.1 $\omega_{\rm sp}$
, transitions into the continuum states are possible, and the quantization of the continuum into discrete box states is likely the source of this dependence. 
Indeed, the energy of some of the peaks in figure \ref{fig:CSVectorVaryBC} show a clear dependence on the size of the simulation sphere.  Of those levels which can be more clearly tracked, a reasonable fit can be obtained of their energy to $1 /R_{\rm grid}^2$.
Most importantly, the general character of the different resonances does not change with the size of the simulation sphere.  In each case that we have examined, we see quantum core plasmons and classical surface plasmons.

The quantum core plasmon discussed above for the simulation size of $R_{\rm grid} = 3.4 R_{\rm MNP}$, occurs at a frequency of 0.7337 $\omega_{\rm sp}$.  
Figure \ref{fig:CSVectorVaryBC} shows that there is consistently a resonance near this frequency regardless of simulation size.  However the peak is sometimes 
accompanied by a series of  levels that are red-shifting with increasing simulation size.

  Repeating the analysis of the previous section on this resonance for a simulation size of $R_{\rm grid} = 4R_{\rm MNP}$ gives results that are nearly identical to those for the quantum core plasmon just discussed for a size of $3.4 \,R_{\rm MNP}$.

Likewise, both of these two simulation sizes have resonances near 0.91 $\omega_{\rm sp}$.  Examining the transitions between Kohn-Sham states for this resonance in the larger simulation volume again yields results very nearly identical to those for the classical surface plasmon just discussed for the smaller simulation volume.  The same shells are involved with similar relative strength between the inversion behavior and the sloshing behavior.  We conclude that the time and frequency dependence of the level occupations for both the quantum core plasmon and classical surface plasmon depend only weakly on simulation size.  This reassures us that although our spectra appear dependent on simulation size, the classification of the dominant resonances is largely independent of simulation size.  We think that the primary size dependence comes from discrete transitions involving empty states far above the Fermi level that are unbound or nearly so.

\subsection{Variation with Particle Size}
We have performed similar analysis of the excitations for smaller MNPs, with 20 and 40 valence electrons.  In each of these systems there is one significant resonance with induced charge at the surface of the particle and a much weaker resonance with induced charge in the core of the particle. 
The surface resonance is about ten times stronger than the core resonance.  

Figures showing the change in electron density, $\Delta P(m,t)$, and the real and imaginary parts of the Fourier transform of $\Delta P(m,t)$ for the two largest resonances for the 20 electron MNP are available as supplementary information on the website of this journal.

Both of these resonances have a significant sloshing response at the driving frequency and a significant inversion response at zero frequency.  All shells seem to have both sloshing and inversion character, consistent with the more mixed character expected for smaller particles.  The most noticable difference between the core and surface plasmon is the addition of transitions from states well below the Fermi level to well above the Fermi level for the core plasmon.  This also occurs for the 100 electron MNP.  The results for the 40 electron MNP (not included) are similar to those for the 20 electron MNP. 

\section{Discussion and Conclusion}

We have seen that any given resonance of a MNP is composed of two types of behavior, which we name sloshing and inversion.
The electrons in the shells nearest the Fermi energy are most able to participate in sloshing behavior.  Sloshing is an oscillation in the occupation of the states in these shells near the Fermi surface, which results in oscillation in the charge density near the surface of the particle.  Inversions, which excite electrons from occupied states (core states or those near the Fermi energy) to unoccupied states, exhibit a monotonic change in the occupation of states involved, and are responsible for oscillation of the charge density in the center of the particle.  In the Fourier transform of $\Delta P(m,t)$, showing the occupation of shells as a function of frequency, sloshing is seen as a $1/\omega$ peak at the driving frequency, while inversions are seen as a $1/\omega$ peak at zero frequency.  Each resonance is ``collective'' to the extent that multiple transitions are involved in any given resonance.  However the resonances with charge density oscillations at the surface (classical surface plasmons) have more states involved in both kinds of transitions.  Sloshing is more significant in these resonances than those resonances with charge density oscillations primarily in the center of the MNP (quantum core plasmons).

Each of the resonances in the spectrum (quantum core plasmons and classical surface plamons) has this 
dual character (sloshing plus inversion).  The dependence of each resonance in the spectrum on simulation size also has a dual character.  The dominant contribution to the main mode depends only weakly on simulation size.  This allows us to give a robust characterization of the mode.  At the same time, each mode mixes with weaker excitations that strongly depend on simulation size.  Since the bound Kohn-Sham states and the ground state electron density have almost no dependence on the size of the simulation sphere, the size dependence should arise mostly from transitions involving highly excited, continuum states.
Our results suggest that we can identify the sloshing behavior as plasmonic and the inversions as more single-particle like. 
We classify individual resonances in the spectra as either classical surface plasmons or quantum core plasmons based on where in the MNP the charge is oscillating.  Each of those resonances, however, displays both sloshing and inversion behavior.  ``Sloshing'' and ``inversion'' indicate how the individual electron states change their population over time. Thus, each of the resonances we examined is a hybridization of plasmonic and single-particle behavior.

This work continues our attempt to understand how the collective behavior of plasmons is built up from individual quantized interacting electrons. 

Having a clear characterization of the MNP plasmons is important for developing a similar understanding for assemblies of MNPs and for hybrid molecules of coupled MNPs and quantum dots.  Surface and quantum core plasmons will have different interparticle coupling due to the different dipole moments they provide.  This must be accounted for in analyzing multiparticle systems.

We have also investigated the dependence of optical absorption spectra on simulation sphere size.  Although the spectra show dependence on simulation size in their details, the broad character of the spectra and those resonances we have examined appear robust with respect to simulation  size.

However for studies of assemblies of MNPs and hybrid molecules, this dependence on simulation size becomes more problematic.  Level crossings are an essential feature of coupled particles.  Additional level crossings that arise from size-dependence effects complicate the analysis of multiparticle systems and are an analytical challenge that must be dealt with.

\section{Acknowledgments} 
The authors wish to thank Eric Shirley for helpful discussions.

\section*{References}
\bibliographystyle{emilysIOP2}
\bibliography{plasmon}

\end{document}